\documentclass[conference]{IEEEtran}
\IEEEoverridecommandlockouts
\usepackage{cite}
\usepackage{amsmath,amssymb,amsfonts}
\usepackage{algorithmic}
\usepackage{graphicx}
\usepackage{textcomp}
\usepackage{xcolor}
\usepackage{verbatim} %注释包
\usepackage{makecell} %table package
\usepackage{multirow}
\newcommand{\tabincell}[2]{\begin{tabular}{@{}#1@{}}#2\end{tabular}}

\def\BibTeX{{\rm B\kern-.05em{\sc i\kern-.025em b}\kern-.08em
    T\kern-.1667em\lower.7ex\hbox{E}\kern-.125emX}}

\makeatletter

\def\ps@IEEEtitlepagestyle{
  \def\@oddfoot{\mycopyrightnotice}
  \def\@evenfoot{}
}
\def\mycopyrightnotice{
  {\footnotesize 978-1-6654-8045-1/22/\$31.00~\copyright~2022 IEEE\hfill} % <--- Change here
  \gdef\mycopyrightnotice{}
}

\def\methodname{AE-smnsMLC}
\def\variantname{AE-smMLC}
\begin{document}

\title{AE-smnsMLC: Multi-Label Classification with Semantic Matching and Negative Label Sampling for Product Attribute Value Extraction
}

\author{\IEEEauthorblockN{Zhongfen Deng}
\IEEEauthorblockA{\textit{Department of Computer Science}\\
\textit{University of Illinois at Chicago}\\
Chicago, Illinois 60607\\
Email: zdeng21@uic.edu}
\and
\IEEEauthorblockN{Wei-Te Chen\\ and Lei Chen}
\IEEEauthorblockA{\textit{Rakuten Institute of Technology}, \\
\textit{Rakuten Group Inc.}\\
Email: weite.chen@rakuten.com\\
lei.a.chen@rakuten.com}
\and
\IEEEauthorblockN{Philip S. Yu}
\IEEEauthorblockA{\textit{Department of Computer Science}\\
\textit{University of Illinois at Chicago}\\
Chicago, Illinois 60607\\
Email: psyu@uic.edu}
}

\maketitle

\begin{abstract}
Product attribute value extraction plays an important role for many real-world applications in e-Commerce such as product search and recommendation.
Previous methods treat it as a sequence labeling task that needs more annotation for position of values in the product text.  
This limits their application to real-world scenario in which only attribute values are weakly-annotated for each product without their position.
Moreover, these methods only use product text (i.e., product title and description) 
and do not consider the semantic 
connection between the multiple attribute values of a given product and its text, which can help attribute value extraction.  
In this paper, we reformulate this task as a multi-label classification task that can be applied for real-world scenario in which only annotation of attribute values is available to train models (i.e., annotation of positional information of attribute values is not available).
We propose a classification model with semantic matching and negative label sampling for attribute value extraction. 
Semantic matching aims to capture semantic interactions between attribute values 
of a given product and its text. 
Negative label sampling  
aims to enhance the model's ability of distinguishing similar values belonging to the same attribute.
Experimental results on three subsets of a large real-world e-Commerce dataset demonstrate the effectiveness and superiority of our proposed model. 
\end{abstract}


\section{Introduction}
\begin{figure}[t]
    \centering
    \includegraphics[scale=0.57]{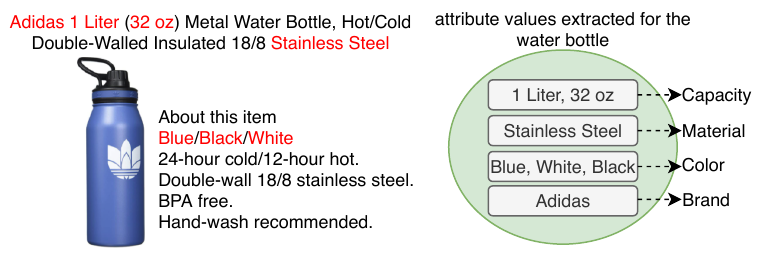}\vspace{-0.1in}
    \caption{An example of water bottle. The left part contains the water bottle's title and description, the right part shows all attribute values extracted for this product.}\vspace{-0.25in}
    \label{fig:intro_example}
\end{figure}

Product attribute value extraction is a fundamental NLP task in e-Commerce which can help improve customer shopping experience. Because it can be utilized for downstream tasks such as product search, product retrieval and recommendation. The most recent state of the art models proposed for this task include \cite{zheng2018opentag,xu2019scaling,karamanolakis2020txtract,yan2021adatag,chen2022extreme}. 
OpenTag \cite{zheng2018opentag} is a sequence labeling model using BiLSTM-CRF with attention mechanism. 
TXtract \cite{karamanolakis2020txtract} makes use of hierarchical taxonomy of categories to help attribute value extraction. AdaTag \cite{yan2021adatag} introduces adaptive decoder for each attribute to share knowledge across different attributes. 
However, all these
models are sequence labeling models requiring annotation of position of attribute values in the product text to train them. In other words, they cannot be trained and used in these cases where annotation of position of values is not available. The above requirement introduces more workload of human annotation than just annotating the existence of attribute values. 
Moreover, in some real-world scenario, the products in a shopping platform or website are already weakly annotated with their attribute values by the merchants when they are created.
Besides, above models   
ignore the semantic connection between attribute values of a product and its text (i.e., title and description). 
In other words, they do not consider the fact that all existing attribute values of a given product come from its
title and description. 
This fact means that for a given product, the semantic meaning of all its existing attribute values can represent this unique product.  
For example, as shown in Fig. \ref{fig:intro_example}, all the attribute values of the product water bottle come from its title and description (red part), each of these attribute values is a short piece of its text. Intuitively, besides represented by the semantic meaning of its title and description, the product of water bottle can also be uniquely represented by the combination of all its attribute values. In others words, the combined semantic meaning of all its attribute values should be as close as possible to the semantic meaning of its text (i.e., title and description).

Based on above intuition, we reformulate product attribute value extraction as a multi-label text classification task which needs less human annotation as mentioned before and can help to  
make a direct semantic connection between attribute values of a product and its text. %information 
The multi-label classification method takes the title and description of a given product as input to predict multiple labels (i.e., attribute values) for this product.

HTCInfoMax \cite{deng2021htcinfomax} achieves good performance for multi-label text classification which  
is designed to predict multiple categories of a given piece of text such as news and scientific paper abstract. Different from that the attribute values of a given product are exactly part of its title and description, the categories of news or scientific paper abstract are summarization of the given text. Thus HTCInfoMax cannot be well suited, although can be applied, for attribute value extraction.

To address the limitation of HTCInfoMax for attribute value extraction, we propose a multi-label classification model with semantic matching and negative label sampling for attribute value extraction called {\methodname}. Specifically, first, similar to HTCInfoMax, besides the text encoder, {\methodname} introduces a label encoder to learn the semantic meaning of attribute values by taking their texts as input. Second, a semantic matching module is designed to make a direct connection 
between the text of a given product and its attribute values by pushing the representation of attribute values towards the representation of the product text.
Third, because attribute values belonging to different attribute (e.g., ``1 liter" of attribute ``capacity" and ``black" of attribute ``color") are easier for the model to distinguish compared with attribute values of the same attribute (e.g., ``1 liter" and ``2 liter" of the attribute ``capacity"). Thus a negative label sampling method is devised to improve the model's ability of distinguishing  
similar attribute values belonging to the same attribute. Our code is available at: https://github.com/zhongfendeng/AE-smnsMLC.

Our work's main contributions are as follows:
1) To our best knowledge, this is the first work to design a multi-label classification model ({\methodname}) for
product attribute value extraction which introduces a label encoder to learn the semantic meaning of attribute values using their text.
2) We propose a semantic matching method in {\methodname} to make a direct interaction between the text of a given product and its all attribute values. 
3) We propose a negative label sampling method to help the model distinguish similar attribute values belonging to the same attribute.
4) Extensive experiments on three subsets of a real-world e-Commerce dataset 
demonstrate the effectiveness of our proposed model.

\section{Problem Formulation}
Given the textual title and description of a product denoted as $T = (w_1, w_2,..., w_L)$, $L$ is its length. The model aims at identifying multiple labels (i.e., attribute values) denoted as $Y = (y_1, y_2, ..., y_M)$ for this product, where $M$ is the number of attribute values this product has. Each of the labels in $Y$ has a pre-defined index which can uniquely represent an attribute value. 
Suppose there are total $N$ attribute values (i.e., labels) in a dataset, the multi-label classification method for attribute value extraction aims at predicting multiple labels out of $N$ labels for each product.

\section{Methodology}
\subsection{Overview}
\begin{figure}[h]
    \centering
    \includegraphics[scale=0.58]{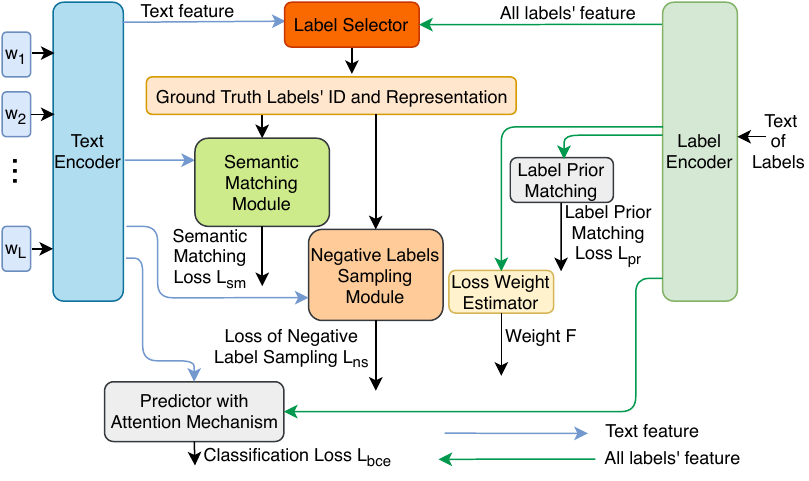}\vspace{-0.1in}
    \caption{The overall architecture of {\methodname}.}\vspace{-0.25in}
    \label{fig:model_architecture}
\end{figure}

As both product text and labels (i.e., attribute values) contain textual information, our proposed model {\methodname} makes use of both product text and label text information. 
Specifically, we design two encoders for encoding the product text and label text information respectively, which are text encoder and label encoder. Then a semantic matching approach with negative label sampling is designed to help both encoders learn better representations for product text and labels. The goal of semantic matching is to push the label feature of a given product towards its text feature which contains richer and complete information about the product. The overall architecture of our proposed method is shown in Fig. \ref{fig:model_architecture}. The major components of our model includes text encoder, label encoder, semantic matching module, negative label sampling module, label selector and loss weight estimator. 
Other components such as predictor with attention mechanism and label prior matching are kept the same as HTCInfoMax \cite{deng2021htcinfomax}. 

\subsection{Text Encoder and Label Encoder}
We use pre-trained BERT-base model as part of the text encoder which takes the product title and description as input. The output of BERT 
is fed into a convolutional layer
which does a convolution operation along the sequence length and outputs the final text representation for the product denoted as $T_{final} \in \mathbb{R}^{c\times d_h}$, where $c$ is the length of sequence after convolution and $d_h$ is the dimension of hidden states of BERT.

\begin{figure}[h]
    \centering
    \includegraphics[scale=0.62]{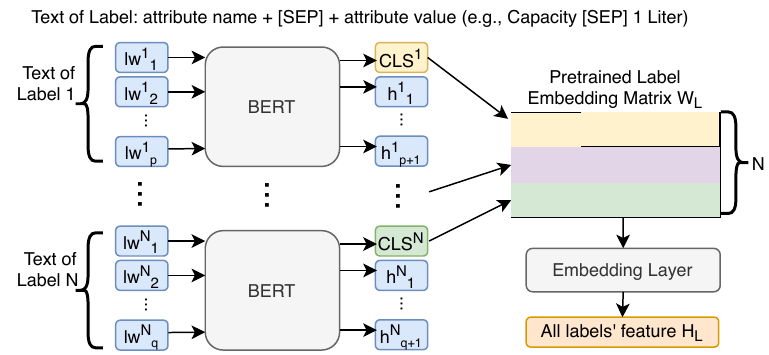}\vspace{-0.1in}
    \caption{The process and structure of label encoder.}\vspace{-0.25in}
    \label{fig:label_encoder}
\end{figure}

Each label has specific text information describing its meaning. It is intuitive to make use of such text information to help learn better representations for labels. Thus, we introduce a label encoder which takes advantage of the label text information to learn the semantic representations $H_L \in \mathbb{R}^{N\times d_l}$ for all labels, $d_l$ is the dimension of label embedding. The process and structure of the label encoder is shown in Fig. \ref{fig:label_encoder}. 

\subsection{Semantic Matching}
Since the product text contains more and complete information about the product, it represents the semantic meaning of the current product well. While the text for each label is relatively short which consists of only several short pieces of text that are attribute name and specific attribute value. However, for each product, the representation learned from its all attribute values (i.e., labels) should also uniquely identify this product, thus we design a semantic matching method to push the labels' representation of a product towards its text feature learned from the complete product text. In this way, the semantic matching can help both text encoder and label encoder learn better text and labels' representation respectively during the training process.
\begin{figure}[t]
    \centering
    \includegraphics[scale=0.60]{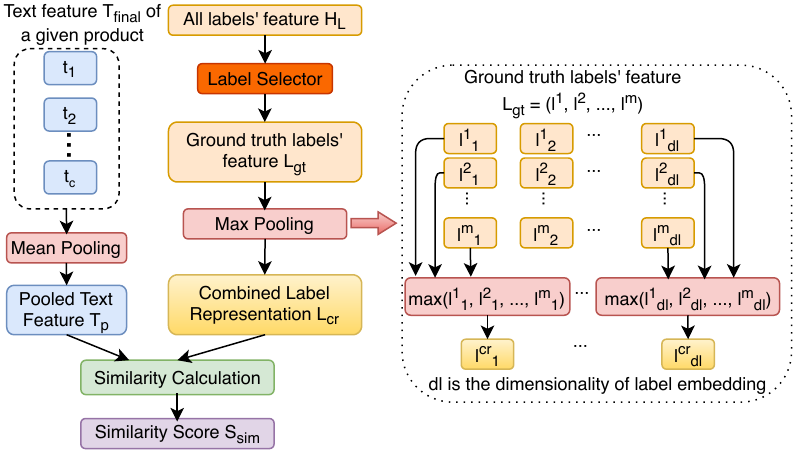}\vspace{-0.1in}
    \caption{The process of semantic matching method.}\vspace{-0.25in}
    \label{fig:semantic_match}
\end{figure}
The details of semantic matching method is shown in Fig. \ref{fig:semantic_match}. The left half shows the overall process of semantic matching, while the right half shows the details of max pooling used in it. 
For the side of text feature, the semantic matching module 
applies mean pooling on $T_{final} \in \mathbb{R}^{c\times d_h}$ 
to obtain pooled text feature $T_p \in \mathbb{R}^{1\times d_h}$.
For the side of label feature, it first uses a label selector to get the ground truth labels' representation $L_{gt} \in \mathbb{R}^{m\times d_l}$ for the current product from $H_L$,
$m$ is the number of ground truth labels of the current product. 
Then max pooling is used to obtain the combined label representation $L_{cr} \in \mathbb{R}^{1\times d_l}$ for the current product as shown in the right half of Fig. \ref{fig:semantic_match}.
After this, 
the semantic matching module calculates the similarity between the combined label embedding and the text feature of the current product, the goal is to push the label embedding as close as possible to the text feature in the feature space which can help the label encoder learn informative representation for all labels. Thus it can help the model enhance the ability of identifying 
multiple attribute values for each product correctly.
We use cosine similarity here for simplicity. The semantic matching loss is
$L_{sm} = -\frac{1}{P} \sum_{i=1}^{P} S_{sim}^{i}.$ 
$P$ is the total number of products (i.e., instances) in the training set.

\begin{figure}[h]
    \centering
    \includegraphics[scale=0.62]{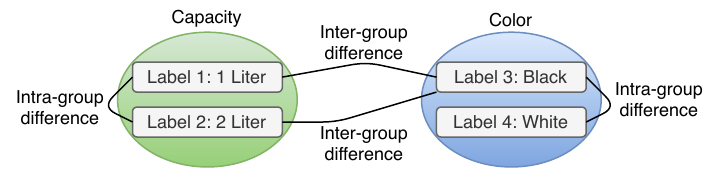}\vspace{-0.1in}
    \caption{An example of labels belonging to two different attribute ``Capacity" and ``Color".}\vspace{-0.25in}
    \label{fig:labels_exmaple}
\end{figure}

\begin{figure}[h]
    \centering
    \includegraphics[scale=0.60]{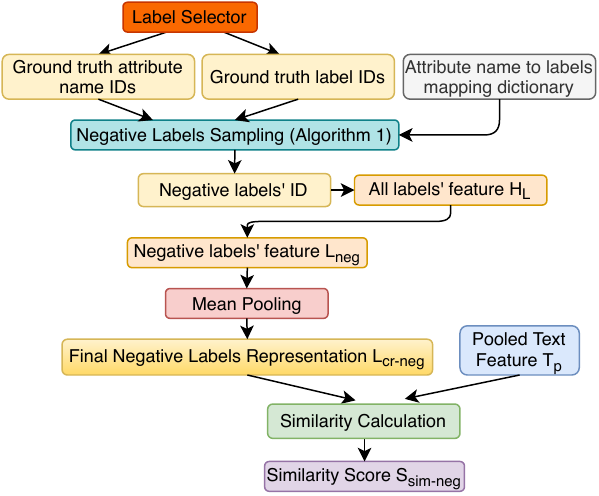}\vspace{-0.1in}
    \caption{The process of negative label sampling.}\vspace{-0.25in}
    \label{fig:negative_labels}
\end{figure}

\subsection{Negative Label Sampling}
The labels (i.e., attribute values) belong to the same attribute may be difficult for models to distinguish because they are much closer to each other than other labels belong to a different attribute. 
Fig. \ref{fig:labels_exmaple} shows such an example. The label "1 Liter" is much closer to the label ``2 Liter" in the same attribute ``Capacity" than to the label ``Black" which belongs to another attribute ``Color". In other words, the difference between labels in the same attribute (i.e., intra-group difference) is smaller than the difference between labels belonging to two different attributes (i.e., inter-group difference).

To help the model enhance the ability of distinguishing similar labels (i.e., attribute values) belonging to the same attribute, 
we propose a negative label sampling method for training the model. The whole process of this method is shown in Fig. \ref{fig:negative_labels} and the specific sampling approach for negative labels is shown in Fig. \ref{fig:algorithm}.
To be specific, we first construct a mapping dictionary $A2L$ which stores all the labels (i.e., attribute values) belong to the same attribute. Then the same label selector as in Fig. \ref{fig:semantic_match} is used to get the ground truth attribute names $AN$ and corresponding labels for each of the attributes $L_{gtIDs}^{AN}$ for the current product. They are taken as input for the negative label sampling algorithm together with the mapping dictionary $A2L$. The output of this sampling algorithm is a list of sampled negative labels for the current product and it is utilized to obtain the negative labels' feature matrix $L_{neg}$ from all labels' feature matrix $H_L$. A mean pooling is applied on the negative labels' feature $L_{neg}$ to get the final compact negative labels' representation $L_{cr-neg}$ for the current product. Similar to the semantic matching module, it calculates the similarity between this compact negative label representation $L_{cr-neg}$ and the pooled text feature $T_p$ of the current product. The goal is to push away the negative labels as far as possible to the current product which can help the model learn more discriminative feature representation for similar labels belong to the same attribute. Thus during the training process, the negative label sampling method aims at minimizing the similarity score $S_{sim-neg}$ between the negative labels and text feature of each product. 
The loss of negative label sampling is 
$L_{ns} = \frac{1}{P} \sum_{i=1}^{P} S_{sim-neg}^{i}$.

\begin{figure}[tb]
\small
%\label{alg:algorithm}
\textbf{Input}: The attribute name to attribute values dictionary $A2L$, ground truth labels' ID list $L_{gtIDs}^{AN}$ and attribute names' ID list $AN$ of current product\\
\textbf{Output}: Negative labels' ID list $NegLabelsID$ for current product
\begin{algorithmic}[1] %[1] enables line numbers
\STATE Let $NegLabelsID=[ ]$.
\FOR{each attribute $A$ in $AN$}
%\WHILE{condition}
\STATE Get ground truth labels $L_{gtIDs}^{A}$ belong to $A$ from $L_{gtIDs}^{AN}$ and the number of labels $L_{gtNum}^{A}$ in $L_{gtIDs}^{A}$.
\STATE Get all labels $L_{All}^{A}$ belong to $A$ from $A2L$.
\STATE Get candidate labels by removing $L_{gtIDs}^{A}$ from $L_{All}^{A}$ and get the number of candidate labels $L_{CandidateNum}$.
\STATE Set number of negative labels $L_{NegNum}=L_{gtNum}^{A}$.
\IF {$L_{CandidateNum} < L_{gtNum}^{A}$}
\STATE Set $L_{NegNum}=L_{CandidateNum}$.
\ENDIF
\STATE Randomly select $L_{NegNum}$ negative labels from candidate labels with equal opportunity for each candidate.%to be selected.
\STATE Add selected negative labels into $NegLabelsID$.
%\ENDWHILE
\ENDFOR
\STATE \textbf{return} $NegLabelsID$
\end{algorithmic}
\caption{Negative labels sampling algorithm}
\label{fig:algorithm}
\end{figure}

\subsection{Final Loss Function}
\subsubsection{Loss of {\methodname}}
As stated before, the semantic matching method and negative label sampling method are proposed to train the model which introduce two losses that are semantic matching loss $L_{sm}$ and loss of negative label sampling $L_{ns}$. The final loss is calculated in Eq. (\ref{eq:final_loss}), where the label prior matching loss $L_{pr}$, loss weight $F \in [0,1]$ and the binary cross entropy loss $L_{bce}$ are inherited from HTCInfoMax.
\begin{equation}\label{eq:final_loss}
\small
    L = L_{bce} + (1-F)\times (L_{sm} + L_{ns}) + F\times L_{pr}.
\end{equation}

\subsubsection{Loss of the Variant {\variantname}} In addition, we also design a variant of our proposed model called {\variantname} which removes the negative label sampling module from {\methodname}. Its final loss 
is $L = L_{bce} + (1-F)\times L_{sm} + F\times L_{pr}$.

\section{Experiment and Analysis}
\subsection{Datasets and Evaluation Metrics}\label{se:dataset}
We use a real-world e-commerce dataset (it is not public due to commercial reasons) possessed by a commercial company 
to conduct experiments for all models.  
All the labels (i.e., attribute values) are annotated by the merchants when they create their products on the company's website. Thus no additional human annotation is needed for training our model.
The text and labels are in Japanese.  
Due to its large size (47 million products), we 
sample products from three different domain as three datasets whose statistics are shown in Table \ref{table:dataset_statistics}. 
Evaluation metrics of multi-label classification including Precision (P), Recall (R), Micro-F1 (MiF1) and Macro-F1 (MaF1) are used to evaluate the performance of models.
\begin{table}[h]
\small
\centering
\caption{Statistics of datasets.  
L and A means the total number of attribute values and attributes in each dataset respectively, AvgL means the average number of attribute values for each product.}
\begin{tabular}{p{1.25cm}<{\centering}| p{2.3cm}<{\centering} p{0.08cm}<{\centering} p{0.18cm}<{\centering} p{0.41cm}<{\centering} p{0.42cm}<{\centering} p{0.35cm}<{\centering} p{0.39cm}<{\centering}} 
 \hline
 Datasets & Domain & A & L & AvgL & Train & Val & Test \\ [0.5ex] 
 \hline
 Dataset 1 & Gardening & 12 & 154 & 5.83 & 4,256 & 69 & 60 \\ [0.5ex]
 Dataset 2 & Plants & 15 & 244 & 4.92 & 4,160 & 219 & 196 \\ [0.5ex]
 Dataset 3 & Gardening Tools & 33 & 218 & 1.24 & 4,160 & 1,254 & 1,268 \\ [0.5ex]
\hline
\end{tabular}
\label{table:dataset_statistics}
\end{table}

\begin{table*}[t]
\small
\centering
\caption{Experimental results of all models on three subsets of different domain.}
\begin{tabular}{p{1.65cm}<{\centering}|p{4.19cm}<{\centering}| p{0.5cm}<{\centering} p{0.5cm}<{\centering} p{0.5cm}<{\centering} p{0.6cm}<{\centering}| p{0.5cm}<{\centering} p{0.5cm}<{\centering} p{0.5cm}<{\centering} p{0.6cm}<{\centering}| p{0.5cm}<{\centering} p{0.5cm}<{\centering} p{0.5cm}<{\centering} p{0.5cm}<{\centering}} 
 \hline
		\multirow{2}{*}{}  & \multirow{2}{*}{\textbf{Models}}    &\multicolumn{4}{c|}{\textbf{Dataset 1 (Gardening)}} & \multicolumn{4}{c|}{\textbf{Dataset 2 (Plants)}} &\multicolumn{4}{c}{\textbf{Dataset 3 (Gardening Tools)}} \\ \cline{3-14} 
		& & \textbf{P} & \textbf{R} & \textbf{MiF1} & \textbf{MaF1} & \textbf{P} & \textbf{R} & \textbf{MiF1} & \textbf{MaF1} & \textbf{P} & \textbf{R} & \textbf{MiF1} & \textbf{MaF1} \\ [0.5ex] 
 \hline
 \multirow{2}{1.0cm}{\rotatebox{0}{\tabincell{c}{BERT-based\\Baselines}}}
 & \tabincell{c}{\textbf{BERT+Linear Layer} \cite{devlin2018bert}} & 80.23 & 39.43 & 52.87 & 17.37 & 73.82 & 26.01 & 38.47 & 11.75 & 83.48 & 47.40 & 60.47 & 15.88 \\ [0.5ex]
 & \tabincell{c}{\textbf{BERT+CNN+Linear Layer} \cite{devlin2018bert}} & 78.50 & 68.86 & 73.36 & 25.61 & 71.23 & 58.24 & 64.08 & 21.07 & 81.64 & 72.12 & 76.58 & 30.41 \\ [0.5ex]
 \hline
 \multirow{3}{1.0cm}{\rotatebox{0}{\tabincell{c}{HTC\\InfoMax\\-based\\Baselines}}}
 & \tabincell{c}{\textbf{HTCInfoMax (TE:BERT)} \cite{deng2021htcinfomax}} & 66.78 & 58.57 & 62.40 & 19.20 & 53.30 & 38.45 & 44.67 & 13.78 & 75.85 & 65.08 & 70.05 & 25.28 \\ [0.5ex]
 \cline{2-14} 
 & \tabincell{c}{\textbf{HTCInfoMax (TE:BERT+CNN)}\\ \cite{deng2021htcinfomax}} & 83.57 & 68.29 & 75.16 & 26.79 & 67.47 & 60.83 & 63.98 & 20.57 & 77.49 & 74.40 & 75.91 & 29.90 \\ [0.5ex]
 \cline{2-14} 
 & \tabincell{c}{\textbf{HTCInfoMax (TE:BERT+CNN)}\\\textbf{-Pretrained CLS Label Emb} \cite{deng2021htcinfomax}} & \textbf{86.48} & 69.43 & 77.02 & 26.61 & \textbf{78.50} & 57.51 & 66.39 & 22.79 & 85.13 & 74.40 & 79.40 & 30.91 \\ [0.5ex]
 \hline
 \multirow{2}{1.0cm}{\rotatebox{0}{\tabincell{c}{Ours}}}
 & \textbf{{\variantname}} & 81.85 & \underline{73.43} & \underline{77.41} & \textbf{30.16} & 78.40 & \underline{57.93} & \underline{66.63} & \underline{23.18} & \underline{82.31} & \textbf{76.36} & \underline{79.22} & \textbf{31.87} \\ [0.5ex]
 & \textbf{{\methodname}} & 81.82 & \textbf{74.57} & \textbf{78.03} & 29.37 & 75.91 & \textbf{62.38} & \textbf{68.49} & \textbf{23.78} & \textbf{85.42} & 76.11 & \textbf{80.50} & 30.41 \\ [0.5ex]
\hline
\end{tabular}
\label{table:results_exp}
\end{table*}

\subsection{Baselines}
Sequence labeling models need labor-intensive and time-consuming human annotation for every token in every product's text (i.e., title and description) which is not available in the stated dataset. Thus no sequence labeling models can be trained on the dataset. Therefore, we only consider classification models as the baselines. Specifically, 
there are two groups of baselines used to compare with our model {\methodname} which are described as follows. Same as in our model, the parameters of the pre-trained BERT-base model used in all these baselines are fixed due to out of GPU memory issue of fine-tuning. The pre-trained BERT is trained with 10 million product descriptions in Japanese.
\subsubsection{BERT-based baselines}
This group uses the same pre-trained BERT-base model \cite{devlin2018bert} as in our model to encode product text (i.e., title and description). 
\textbf{BERT+Linear Layer} feeds the CLS embedding 
outputted by BERT to a linear layer which predicts the product's attribute values.
\textbf{BERT+CNN+Linear Layer} utilizes an additional 
CNN layer on top of BERT.  
The output of CNN is fed to a linear layer to do prediction. 
\subsubsection{HTCInfoMax-based baselines}
To be fairly compared with our model. All HTCInfoMax-based baselines \cite{deng2021htcinfomax} adopt the same pre-trained BERT-base model as text encoder or part of it, and the same label embedding layer as in the label encoder of our model is used as the structure encoder of HTCInfoMax. 
\textbf{HTCInfoMax (TE:BERT)} uses the pre-trained BERT as text encoder, the label embedding layer of the structure encoder is randomly initialized.
\textbf{HTCInfoMax (TE:BERT+CNN)} stacks an additional CNN layer on top of BERT to form the text encoder. 
The label embedding layer of the structure encoder is also randomly initialized.
\textbf{HTCInfoMax (TE:BERT+CNN)-Pretrained CLS Label Emb} has the same text encoder 
and label encoder 
as our model, the pretrained label embedding generated by BERT is used to initialize the label embedding layer.

\subsection{Experimental Setting}
All baselines and our model adopt the same setting for all hyper-parameters. To be specific, the maximum length of token sequence of product text is set to 256, the dimensions of text and label embedding are 768, the kernel size of CNN layer is 4. 
HTCInfoMax-based baselines are re-implemented by us using PyTorch based on the released code \cite{deng2021htcinfomax}. We implement our model and BERT-based baselines using PyTorch.
All experiments are conducted on a single NVIDIA Quadro M6000 GPU sever with 12G GPU memory.

\subsection{Experimental Results and Analysis}
The experimental results of our model and baselines on the three datasets  
are shown in Table \ref{table:results_exp}. 
We can see that our proposed model {\methodname} generally outperforms all baselines on Recall, Micro-F1 and Macro-F1 including the strong baseline in the group of HTCInfoMax-based baselines which utilizes pretrained CLS label embedding to initialize its structure encoder (i.e., the third from last row in Table \ref{table:results_exp}). Although this baseline has higher precision score on Gardening and Plants, our model obtains much better Recall, Micro-F1 and Macro-F1 scores by improvements of 7.40\%, 1.31\%, 10.37\% on Gardening and by improvements of 8.47\%, 3.16\%, 4.34\% on Plants respectively. 
This demonstrates that the semantic matching and negative label sampling methods help our model perform better on all labels, because Micro-F1 and Macro-F1 scores are more reliable metrics that evaluate models in a much more comprehensive way. Furthermore, our model outperforms the previously mentioned strongest baseline on Precision, Recall and Micro-F1 on Gradening Tools dataset by improvements of 0.34\%, 2.30\% and 1.39\% respectively, which also verifies the superiority of our model.
In addition, the consistent performance of our model across three different datasets demonstrates that our proposed model can learn better text representation and better representations for all attribute values (i.e, labels) by matching the semantic representation of all %combined
ground truth attribute values of a given product towards its text representation. In other words, the performance consistency validates the effectiveness of our proposed semantic matching method. 

Besides, from the comparison between models in the group of HTCInfoMax-based baselines, we can see that pretrained label embedding generated by using the text (e.g., ``stainless steel") of attribute values can help improve the performance. This indicates that label text information is helpful for product attribute value extraction in terms of reformulating it as a multi-label classification task.
In addition, the baselines with CNN layer performing much better than that without CNN indicates the importance of CNN layer in the text encoder.

\subsection{Ablation Study}
We conduct ablation studies to demonstrate the effect  
of different components in our model such as negative label sampling and pooling method. 
We also design a third variant of our model called ``{\methodname} w/o LabelPrior" which removes the label prior matching module for more ablation study. The results of ablation studies are shown in Table \ref{table:results_ablation}. For each of the three models in this table, two pooling methods for ground truth labels in semantic matching module are experimented. One is max pooling as shown in Fig. \ref{fig:semantic_match}, the other is mean pooling which replaces the max pooling. 
\begin{table}[t]
\small
\centering
\caption{Results of ablation studies on Dataset 1.}
\begin{tabular}{p{2.0cm}<{\centering}|p{1.45cm}<{\centering}| p{0.5cm}<{\centering} p{0.5cm}<{\centering} p{0.5cm}<{\centering} p{0.6cm}<{\centering}} 
 \hline
	\multirow{1}{*}{\textbf{Models}}  & \multirow{1}{*}{\tabincell{c}{\textbf{SM Pooling}}}    
	& \textbf{P} & \textbf{R} & \textbf{MiF1} & \textbf{MaF1} \\ [0.5ex] 
 \hline
 \multirow{2}{2.0cm}{\rotatebox{0}{\tabincell{c}{\textbf{{\methodname}} \\ \textbf{w/o LabelPrior}}}}
 & \tabincell{c}{Mean Pool} & \textbf{86.10} & 72.57 & \textbf{78.76} & 28.91 \\ [0.5ex]
 & \tabincell{c}{Max Pool} & 79.94 & 71.71 & 75.60 & 27.55 \\ [0.5ex]
 \hline
 \multirow{2}{2.0cm}{\rotatebox{0}{\tabincell{c}{\textbf{{\variantname}}}}}
 & \tabincell{c}{Mean Pool} & 76.07 & 70.86 & 73.37 & 26.96 \\ [0.5ex]
 %\cline{2-6} 
 & \tabincell{c}{Max Pool} & \textbf{\textit{81.85}} & 73.43 & 77.41 & \textbf{30.16} \\ [0.5ex]
 \hline
 \multirow{2}{2.0cm}{\rotatebox{0}{\tabincell{c}{\textbf{{\methodname}}}}}
 & Mean Pool & 79.17 & \textbf{76.00} & 77.55 & 27.28 \\ [0.5ex]
 & Max Pool & 81.82 & \textbf{\textit{74.57}} & \textbf{\textit{78.03}} & \textbf{\textit{29.37}} \\ [0.5ex]
\hline
\end{tabular}
\label{table:results_ablation}
\end{table}

From Table \ref{table:results_ablation}, one can see that max pooling is generally performing better than meaning pooling. And our models with (the last two rows) or without (the first two rows) label prior matching achieve comparably similar results, which verifies the effectiveness of our proposed model in capturing the semantic connections between attribute values of each product and its text. It also indicates that the label prior matching helps. Furthermore, the comparison between {\methodname} and {\variantname} in Table \ref{table:results_ablation} (the last four rows) and Table \ref{table:results_exp} (the last two rows) shows that {\methodname} generally performs better than {\variantname} without negative label sampling. This demonstrates that negative label sampling can help the model distinguish similar labels  
by learning more discriminative feature for them and thus improves the performance.

\subsection{Case Study}
\begin{figure}[h]
    \centering
    \includegraphics[scale=0.43]{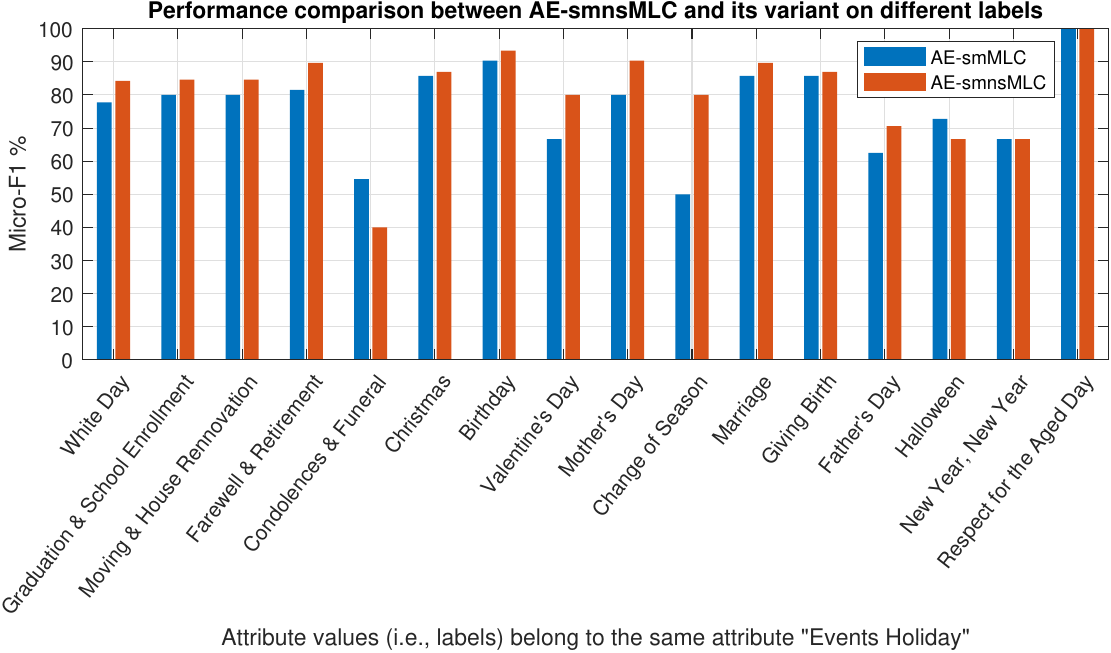}\vspace{-0.1in}
    \caption{The case study for negative label sampling.}\vspace{-0.25in}
    \label{fig:case_study}
\end{figure}

To further verify the usefulness of negative label sampling, we conduct a case study on Gardening dataset.  
Specifically, we select the prediction results of {\methodname} and its variant {\variantname} on all labels belong to some attribute (e.g., ``Events/Holiday") from Gardening dataset and calculate Micro-F1 scores for these labels. The comparison between the performance results of our model and its variant without negative label sampling on all labels belong to the attribute ``Events/Holiday" is shown in Fig. \ref{fig:case_study}. One can see that {\methodname} performs better on almost all labels belong to this attribute which demonstrates that negative label sampling can help the model learn much more discriminative feature for labels belong to the same attribute and thus improve the model's ability of distinguishing similar labels.

\section{Related Work}
\subsection{Product Attribute Value Extraction and Multi-Label Text Classification}
Earlier models for product attribute value extraction are rule-based extraction methods \cite{gopalakrishnan2012matching,vandic2012faceted}. 
Later on, \cite{more2016attribute} treats this task as a sequence labeling task and many neural network based sequence labeling models with different techniques such as attention mechanism, using hierarchical taxonomy of product, adaptive CRF decoder and so on are designed \cite{zheng2018opentag,xu2019scaling,mehta2021latex,karamanolakis2020txtract,yan2021adatag}.
Most of multi-label text classification models can be categorized into two groups. One group is local models which 
build a classifier for each label or labels in the same 
level of the label taxonomy \cite{wehrmann2018hierarchical,banerjee2019hierarchical,huang2019hierarchical}.
The other group is global models building one classifier for all labels \cite{johnson2015effective,mao2019hierarchical,wu2019learning,peng2018large,peng2019hierarchical,zhou2020hierarchy,deng2021htcinfomax}. The latter four models make use of label structure information. In addition, attention-based models are popular for text classification in recent years \cite{you2019attentionxml,chang2020taming,deng-etal-2020-hierarchical}. 

\section{Conclusion}
We propose a multi-label classification model for product attribute value extraction which can be applied for real-world scenario in which only attribute values are annotated without their position information in the product text. Our proposed model introduces a label encoder, 
a semantic matching  
and a negative label sampling method. Semantic matching aims to capture the semantic interactions between attribute values of a given product and its text. 
Negative label sampling helps enhance the model's ability to distinguish similar labels.
Experimental results on three subsets of a large-size real-world e-commerce dataset demonstrates the superiority of our model for 
product attribute value extraction. 

\section*{Acknowledgment}

We thank the reviewers for their comments and feedback. This work is supported in part by NSF under grants III-1763325, III-1909323,  III-2106758, and SaTC-1930941.

%\section*{References}

\vspace{12pt}

\end{document}